# Why Lead Methylammonium tri-IODIDE perovskite-based solar cells requires a mesoporous electron transporting scaffold (but not necessarily a hole conductor)

*Eran Edri[1], Saar Kirmayer[1], Alex Henning[2], Sabyasachi Mukhopadhyay[1], Konstantin Gartsman[3], Yossi Rosenwaks[2], Gary Hodes[1], David Cahen[1,\*]*

[1]Department of Materials and Interfaces, Faculty of Chemistry Weizmann Institute of Science, 76100, Rehovot, Israel.

[2]Department of Electrical Engineering–Physical Electronics, Faculty of Engineering, Tel Aviv University, Ramat-Aviv, 69978, Israel.

[3]Department of Chemical Research Support, Faculty of Chemistry Weizmann Institute of Science, 76100, Rehovot, Israel.

**Abstract.** $CH_3NH_3PbI_3$-based solar cells were characterized with electron beam-induced current (EBIC), and compared to $CH_3NH_3PbI_{3-x}Cl_x$ ones. A spatial map of charge separation efficiency in working cells shows p-i-n structures for both thin film cells. Effective diffusion lengths, $L_D$, (from EBIC profile) show that holes are extracted significantly more efficiently than electrons in $CH_3NH_3PbI_3$, explaining why $CH_3NH_3PbI_3$-based cells require mesoporous electron conductors, while $CH_3NH_3PbI_{3-x}Cl_x$ ones, where $L_D$ values are comparable for both charge types, do not.

Finding a way for making efficient yet inexpensive photovoltaic devices is a key step towards enhancing PV technology incorporation in tomorrow's energy mix.[1] Perovskite organic-inorganic lead iodide-based solar cells have emerged as a new and promising class of materials for photovoltaic applications with efficiencies reaching 10% nearly from the start.[2,3] Also because of the diversity in cell types the question how these cells accomplish this, is a burning one. Right from the early reports on >9% efficient cells some major differences between cell types appeared. One concerns whether chlorine is used in the absorber precursor solution or not, with the products denoted $CH_3NH_3PbI_{3-x}Cl_x$ [3] or $CH_3NH_3PbI_3$.[2,4] Another one concerns the type of mesoporous (mp) layer used: one that can drain the electrons from the absorber (e.g. mp-$TiO_2$ or ZnO)[5,6], or one that is electronically "inert" and cannot accept electrons from the absorber (e.g. mp-$Al_2O_3$ or $ZrO_2$)[3,7,8]. Although the band gap is not much affected, Cl incorporation was reported to increase the diffusion length of charge carriers by an order of magnitude,[9] from ca. 130 nm to > 1000 nm. While the reason for this remains to be clarified, this quality was used to explain why $CH_3NH_3PbI_3$, if used as an absorber in a flat design (i.e., without mesoporous scaffold), has efficiencies of 2-3%, while with $CH_3NH_3PbI_{3-x}Cl_x$ nearly 16% can be reached.[10] However, by careful chemical interplay, one that leads to efficient surface coverage and obviates the need for long diffusion lengths of the electrons, a cell of > 15% efficiency can be obtained with $CH_3NH_3PbI_3$ and mp-$TiO_2$.[11]

Furthermore, in a certain case, using $CH_3NH_3PbI_3$, a hole conductor is not required at all and > 8% efficient device can be obtained (with mp-$TiO_2$ as electron conductor and the Perovskite absorber as the hole conductor).[12,13]

It was suggested that at least the $CH_3NH_3PbI_{3-x}Cl_x$ – based cells may be of the p-i-n type.[3] The undisputed high crystallinity of the perovskite layers indicates, though, that there must be a difference with the most common of p-i-n cells, the a-Si – based one. Indeed, we showed here that, instead, the cell that resembles the $CH_3NH_3PbI_{3-x}Cl_x$ – based one the most is a p-AlGaAs / i-MQW GaAs.AlGaAs / n-AlGaAs one,[15] with the perovskite in the role of the MQW. Given that the perovskite materials are prepared at close to room temperature from (organic) solution and the high level of sophistication and exquisite crystal quality of the III-V-based materials, this analogy is astounding.

Having similar energy band gap valu, and similar ionization energy, this begs the questions what type of a cell is the $CH_3NH_3PbI_3$ material suitable for, and why.

Using electron beam-induced current (EBIC) we have recently shown that the $CH_3NH_3PbI_{3-x}Cl_x$ based cells, which (if cast from solution) function best with an inert mp-scaffold, operate as a p-i-n device.[14] The two halves of the junction (p-i and i-n) work in conjunction to give the high $V_{OC}/E_{gap}$ that characterizes these cells. In a p-i-n device (in contrast to a p-n junction, for example), the built-in electric field in the nearly intrinsic absorber is induced by the two selective contacts. Given that the absorber material is of high enough electronic quality and the effective diffusion length of both charges is comparable, this field supports charge separation throughout the whole absorber layer.[16] At the same time, the energy offsets between the conduction (valence) band edge of the absorber-electron transport material and the valence (conduction) band of the absorber-hole transport material act to selectively transfer one type of charge carrier. Furthermore, we concluded that *also when an inert scaffold is used*, the same photovoltaic mechanism pertains. The EBIC measurement provided a rather direct way to obtain the effective diffusion lengths of the charge carriers in the $CH_3NH_3PbI_{3-x}Cl_x$ cells (1.9 ± 0.2 μm for electrons and 1.2 ± 0.1 μm for holes, in reasonable agreement with what was deduced from modeling time-resolved optical measurements on partial devices[9]). Here we focus on devices based on $CH_3NH_3PbI_3$ and by using similar tools as before show that despite the differences in material properties, which brings the requirement for the use of a mesoporous electron transport layer, the p-i-n mode of operation also underlies these cells.

In an EBIC experiment, the electron beam of a scanning electron microscope is used to generate multiple electron-hole pairs, which in analogy with *photo*-generated electron-hole pairs, when subjected to the device internal charge separating driving force, produce the *electron*-generated current.[17] This current is measured externally, and since it is correlated with the scanning beam position, an image of the locus of charge separation is produced, with the peaks in intensity showing the location(s) of the highest charge collection efficiency, i.e., where the electronic junction is and how well the charge carriers are collected. This is illustrated in Figure 1, with the layer-widths exaggerated for clarity.

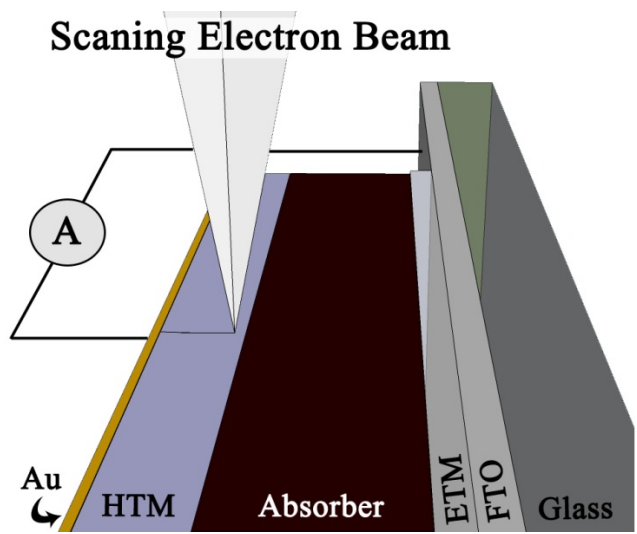

**Figure 1.** A schematic illustration of an electron beam induced current (EBIC) on a cross-section of a planar perovskite solar cell. All parts in the image are not to scale.

In Figure 2 the Secondary electron (SE) and EBIC images of $CH_3NH_3PbI_{3-x}Cl_x$ and $CH_3NH_3PbI_3$ planar solar cells are presented with the corresponding line scans, showing the two distinctively different EBIC profiles of ('flat') cells, that apart from the difference in the composition of the precursor solutions, are made in a similar way.

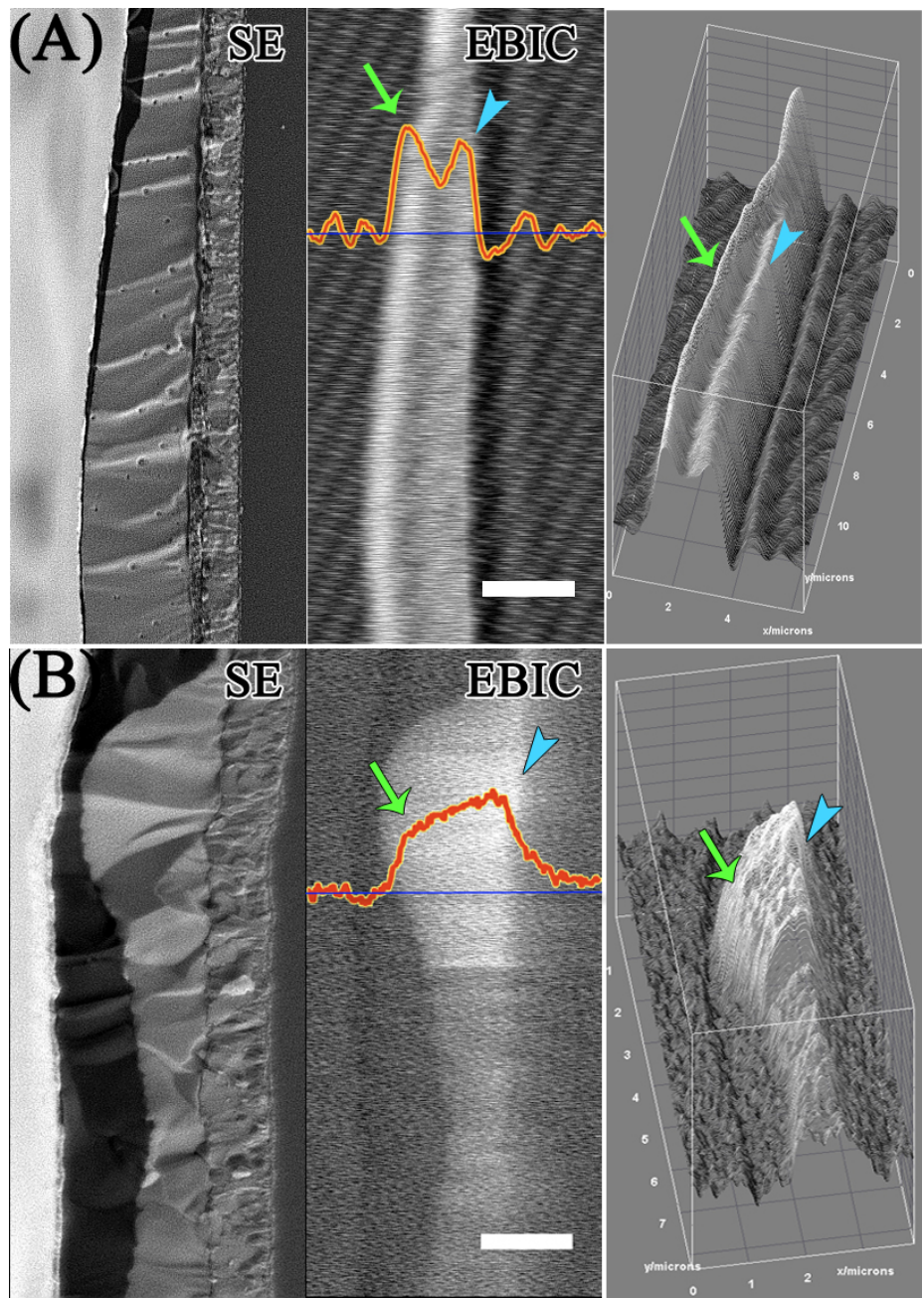

**Figure 2.** Secondary electron (SE) and electron beam induced current (EBIC) images of cross sections of $CH_3NH_3PbI_{3-x}Cl_x$ **(A),** scale bar is 2 μm, and $CH_3NH_3PbI_3$ **(B)** scale bar is 1 μm, planar solar cells. Line scans were taken at the lines positions. The arrows show the peaks for the I-Cl and where the peaks are in the case of the pure iodide. The right panel is a 3D surface plot of the EBIC images. The ripples observed in the EBIC image of **(A)** outside the (semi-) conductive regions are due to background noise.

The EBIC profile of the $CH_3NH_3PbI_{3-x}Cl_x$-based solar cell (Figure 2A) shows current generation throughout the absorber layer (but not in the selective contact layers), together with the two-peak pattern *in* the absorber, close to the selective contacts. This EBIC line profile is characteristic of a p-i-n photovoltaic device, where the intrinsic layer is of high electronic quality material. Although charge is being separated throughout the absorber layer, the two peaks, which are not symmetric in their profile, represent the two prime locations for charge separation in the device: the p-i (indicated by a green arrow) and i-n (half) junctions (indicated by a blue pointer). As can be seen, the "p-i" peak is higher than the "i-n" one throughout the whole film. The line profile of a device without HTM does not have this additional peak close to the HTM.[14] There, a heterojunction is formed between the ETM and the absorber, which is expressed as a single EBIC peak close to the junction.[14, 15] In the analogous solar cell, using pristine $CH_3NH_3PbI_3$ as the absorber (Figure 2B), we also find a line profile that corresponds to a p-i-n device. As above, there is current generation throughout the absorber layer, with two peaks, each near a selective contact. However, now there is a big difference between the intensity of the two peaks, with the dominant one closer to the ETM and a smaller one, appearing as a shoulder to the left of the main peak, close to the HTM. This is more clearly observed in the 3D surface plot (Figure 2B, right panel). The peak close to the compact $TiO_2$ is high enough so that it conceals the peak close to the HTM.

The effective diffusion lengths can be extracted from the EBIC signal by fitting a decay curve to the decay of the intensity peak as was previously shown for the $CH_3NH_3PbI_{3-x}Cl_x$ system. As the EBIC signal intensity is a manifestation of the charge extraction efficiency, the ratio between the peaks is an indication of the ratio between the effective diffusion lengths. Here the effective diffusion length of the holes appears to be much longer than that of electrons, resulting in one peak much higher than the other, to such an extent that the one close to the ETM is partly obscuring the peak closer to the HTM.

Another difference between the two systems, which can be seen in the EBIC images, is that while the $CH_3NH_3PbI_{3-x}Cl_x$ forms a uniform EBIC pattern throughout the film. $CH_3NH_3PbI_3$ forms a distribution of intensities in different grains and areas with stronger intensities in some grains than in others. For example, the part of the cross-sectioned crystal shown in Figure 2B, which is wider than 1 µm shows an EBIC signal throughout the crystal cross section, indicating that charges generated close to either of the selective contacts, can still reach the other contact before recombining. This finding is incompatible with a diffusion length of ca. 100 nm, as deduced earlier,[9,18] because then we would not see a significant EBIC signal from regions away from a contact by more than a few multiples of the diffusion length. From this we conclude that at least in parts of the $CH_3NH_3PbI_3$ film, the effective diffusion length of electrons is a good fraction of 1 µm. This discrepancy can be explained at least in part by the different method of measurement. In EBIC experiments, we measure a complete device structure, rather than a partial one, and so the effective diffusion length (i.e., one that considers the effect of electric fields, induced by *both* selective contacts in the ambipolar absorber) is observed *directly*. In addition, in EBIC the quasi-free carriers rather than excitons are measured.

While the effect of the presence of $Cl^-$ and excess $CH_3NH_3^+$ in the precursor solution on the materials structure and properties is yet to be fully understood, part of the collective effect of their presence is that for $CH_3NH_3PbI_{3-x}Cl_x$ we mostly find films made of large grains, up to a size of a few microns, while for $CH_3NH_3PbI_3$ we mostly find films with grain sizes of a 200-400 nm. In the EBIC images of $CH_3NH_3PbI_{3-x}Cl_x$ grains appear to produce current uniformly throughout the grain while in the EBIC images of multi-crystalline $CH_3NH_3PbI_3$, film each grain contributes differently to the current, as can be seen by the different EBIC contrast of each of the grains (Figure 3). The pointers in Figure 3A indicate two adjacent grains with different EBIC contrast. Furthermore, the granular morphology as well the different EBIC contrast of each grain pertain if mesoporous $TiO_2$ is used as the substrate with and without HTM (Figure 3B and C; the corresponding I-V curves of the three different devices are shown in Figure S2).

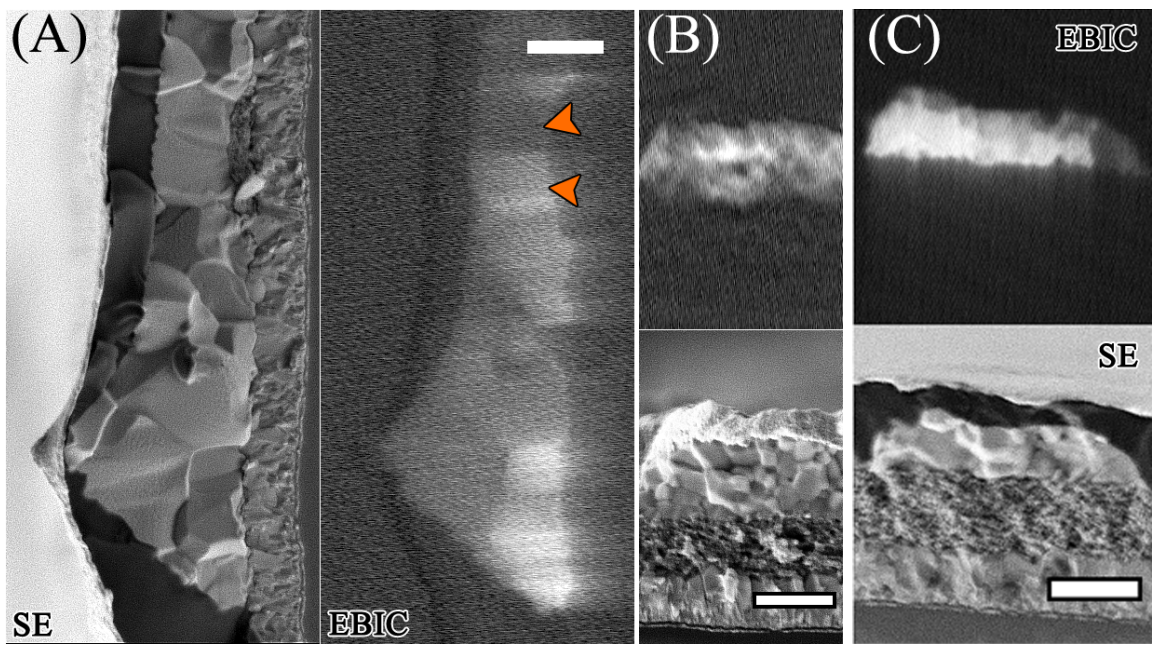

**Figure 3.** Secondary electron and Electron beam induced current images of granular films. (A) Au\HTM\$CH_3NH_3PbI_3$\compact $TiO_2$\FTO. (B) Au\$CH_3NH_3PbI_3$\mesoporous $TiO_2$\compact $TiO_2$\FTO. (C) Au\HTM\$CH_3NH_3PbI_3$\mesoporous $TiO_2$\compact $TiO_2$\FTO. The EBIC image on the right side of (A) shows the different contribution of each grain to the current collected, indicated by the different EBIC contrast. The two pointers indicate two neighboring grains, with similar thickness and morphology but different EBIC contrast. The EBIC image in (B) and (C) shows granular features smaller than in (A) but similar EBIC features. Scale bar is 1 μm.

The EBIC pattern observed in the granular films of $CH_3NH_3PbI_3$ resembles that of certain CIGS solar cells. There, differences in orientation of the grains was invoked to explain the different EBIC contrast from each grain.[19,20] X-ray diffraction (supplementary information Figure S1) of $CH_3NH_3PbI_3$ and $CH_3NH_3PbI_{3-x}Cl_x$ films shows that the latter seem to have stronger orientation if spin-coated on the substrate. Clearly, though, apart from the quest for ever-higher efficiencies, much more research is required on these and other materials-related issues. We also find that the EBIC contrast from the grain boundaries of multi-crystalline $CH_3NH_3PbI_3$ sample (Figure 3) is not different from that of the grains themselves, indicating negligible charge separation and recombination takes place at the grain boundaries. However, Kelvin probe force microscopy (KPFM) of perovksite films deposited on ITO (Figure 4) reveals a few tens of mV (~ 40 mV) surface potential difference between the grain boundary and grain bulk, with a higher contact potential difference (CPD), i.e., a higher work function, at the GB. This indicates the presence of a small potential barrier for electron transport between grains due to some upward bending of the conduction band, as will result from electron depletion at the grain boundaries. Under illumination, this shallow barrier is reduced, so that it has a negligible effect on the photovoltaic performance. Cross-section measurements (Figure 4 bottom panel, right) show a higher CPD at grain boundaries, consistent with the top plane measurements on the perovskite film.

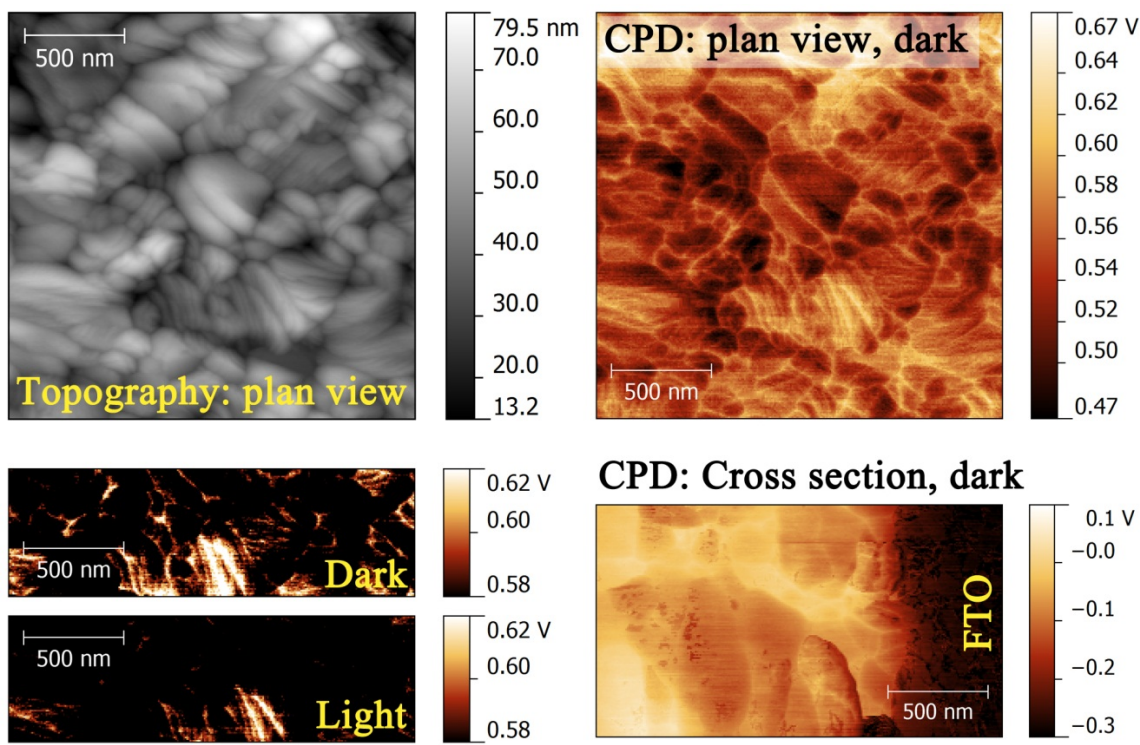

**Figure 4.** KPFM images of $CH_3NH_3PbI_3$ on ITO. Topography and contact potential difference (CPD) images of plan view scans are shown in the upper panel. Bottom panel: *left* - plan view scans in the dark and under illumination; *right* - CPD of a cross-section (with $TiO_2$\FTO contact).

To conclude, EBIC imaging of cross sections of 'flat' cells show that $CH_3NH_3PbI_3$-based cells can, similarly to the $CH_3NH_3PbI_{3-x}Cl_x$ ones, operate as p-i-n devices. However the difference in EBIC contrast near each of the selective contacts indicates that the effective diffusion length of electrons is shorter than that of holes ($L_{eff, e-}/L_{eff, h+} < 1$), with the latter being at least 1 μm. We have also shown that the tri-iodide material, if spin casted from solution forms (mostly) a multi-crystalline layer with average grain size of a few hundreds of nanometers with each crystal grain contributing differently to the external (electron-generated) current. CPD measurements indicate that only a very shallow barrier, due to upward conduction band bending, exists for electrons to cross the grain boundaries. These observations explain why $CH_3NH_3PbI_3$ requires a mesoporous $TiO_2$ to make an efficient solar cell, but not necessarily a hole conductor. While holes can travel a long distance before recombining, electrons can travel only a relatively short distance and therefore require a separate conduit to the front electrode. These impediments are overcome by the use of the mesoporous $TiO_2$, so that the electrons need to cross only a short distance before being injected to the $TiO_2$ and the granularity-directionality of the film becomes less important.

Whether the difference between the tri-iodide material and its chlorine incorporated counterpart is a result of e.g., different internal electric fields, crystal orientation or doping (without ruling out other possible reasons) requires more research.

Finally, because a mesoporous electron transporter is required for these pure iodide-based cells to work efficiently, we suggest that the path for further improving performance of the tri-iodide based PV system, is to consider it as an extremely thin absorber, rather than a thin film solar cell.

ACKNOWLEDGMENT

We thank the Leona M. and Harry B. Helmsley Charitable Trust, the Israel Ministry of Science's "Tashtiot" program, the Israel National Nano-Initiative's Focused Technology Area program, the Nancy and Stephen Grand Center for Sensors and Security and the Weizmann-UK Joint Research Program, for support. D.C. holds the Sylvia and Rowland Schaefer Chair in Energy Research.

REFERENCES

(1) *Fundamentals of materials for energy and environmental sustainability*; Ginley, D. S.; Cahen, D., Eds.; 1st ed.; Cambridge University Press: Cambridge; New York, 2012.

(2) Kim, H.-S.; Lee, C.-R.; Im, J.-H.; Lee, K.-B.; Moehl, T.; Marchioro, A.; Moon, S.-J.; Humphry-Baker, R.; Yum, J.-H.; Moser, J. E.; Grätzel, M.; Park, N.-G. *Sci. Rep.* **2012**, *2*.

(3) Lee, M. M.; Teuscher, J.; Miyasaka, T.; Murakami, T. N.; Snaith, H. J. *Science* **2012**.

(4) Kojima, A.; Teshima, K.; Shirai, Y.; Miyasaka, T. *J Am Chem Soc* **2009**, *131*, 6050–6051.

(5) Bi, D.; Boschloo, G.; Schwarzmüller, S.; Yang, L.; Johansson, E. M. J.; Hagfeldt, A. *Nanoscale* **2013**.

(6) Kim, H. S.; Lee, C. R.; Im, J. H.; Lee, K. B.; Moehl, T.; Marchioro, A.; Moon, S. J.; Humphry-Baker, R.; Yum, J. H.; Moser, J. E.; Gratzel, M.; Park, N. G. *Sci Rep* **2012**, *2*.

(7) Bi, D.; Häggman, L.; Boschloo, G.; Yang, L.; Johansson, E. M. J.; Hagfeldt, A. *RSC Adv.* **2013**.

(8) Ball, J. M.; Lee, M. M.; Hey, A.; Snaith, H. *Energy Environ. Sci.* **2013**.

(9) Stranks, S. D.; Eperon, G. E.; Grancini, G.; Menelaou, C.; Alcocer, M. J.; Leijtens, T.; Herz, L. M.; Petrozza, A.; Snaith, H. J. *Science* **2013**, *342*, 341–344.

(10) Docampo, P.; Ball, J. M.; Darwich, M.; Eperon, G. E.; Snaith, H. J. *Nat. Commun.* **2013**, *4*.

(11) Burschka, J.; Pellet, N.; Moon, S.-J.; Humphry-Baker, R.; Gao, P.; Nazeeruddin, M. K.; Grätzel, M. *Nature* **2013**, *499*, 316–319.

(12) Etgar, L.; Gao, P.; Xue, Z. S.; Peng, Q.; Chandiran, A. K.; Liu, B.; Nazeeruddin, M. K.; Gratzel, M. *J Am Chem Soc* **2012**, *134*, 17396–17399.

(13) Laban, W. A.; Etgar, L. *Energy Environ. Sci.* **2013**.

(14) Edri, E.; Kirmayer, S.; Mukhopadhyay, Sabayasachi; Gartsman, Konstantin; Hodes, Gary; Cahen, David. *Submitted*.

(15) Araújo, D.; Romero, M. J.; Morier-Genoud, F.; García, R. *Mater. Sci. Eng. B* **1999**, *66*, 151–156.

(16) Fonash, S. J. *Solar cell device physics*; Academic Press/Elsevier: Burlington, MA, 2010.

(17) Holt, D. B.; Joy, D. C. *SEM microcharacterization of semiconductors*; Academic, 1989.

(18) Xing, G.; Mathews, N.; Sun, S.; Lim, S. S.; Lam, Y. M.; Grätzel, M.; Mhaisalkar, S.; Sum, T. C. *Science* **2013**, *342*, 344–347.

(19) Kawamura, M.; Yamada, T.; Suyama, N.; Yamada, A.; Konagai, M. *Jpn. J. Appl. Phys.* **2010**, *49*, 062301.

(20) Minemoto, T.; Wakisaka, Y.; Takakura, H. *Jpn. J. Appl. Phys.* **2011**, *50*, 031203.

(21) Ito, S.; Chen, P.; Comte, P.; Nazeeruddin, M. K.; Liska, P.; Péchy, P.; Grätzel, M. *Prog. Photovolt. Res. Appl.* **2007**, *15*, 603–612.

**Materials and Methods.**

*Device Fabrication:* F-doped tin oxide, FTO, transparent conducting substrates (Pilkington TEC15) were cut and cleaned by sequential 15 min sonication in warm aqueous alconox solution, deionized water, acetone and ethanol, followed by drying in a $N_2$ stream. A compact ca. 100 nm thin $TiO_2$ was then applied to the clean substrate by spray pyrolysis of 20 mM titanium diisopropoxide bis(acetylacetonate) solution in isopropanol using air as carrier gas on a hot plate set to 350°C, followed by annealing at 500°C for 1 h in air. The mesoporous layer was applied by adapting the procedure from the literature[21] to spin coating by diluting in ethanol to the needed viscosity, followed by spin coating at 3000 RPM for 45 seconds.

A $CH_3NH_3PbI_3$ solution was prepared as described elsewhere. In short, $CH_3NH_3I$ was prepared by mixing methyl amine (40% in methanol) with hydroiodic acid (57% in water; CAUTION: exothermic reaction) in a 1:1 molar ratio in a 100 mL flask under continuous stirring at 0°C for 2 hr. $CH_3NH_3I$ was then crystallized by removing the solvent in a rotary evaporator, washing three times in diethyl ether for 30 min and filtering the precipitate. The material, in the form of white crystals, was then dried overnight in vacuum at 60°C. It was then kept in a dark, dry environment until further use. A 40%wt solution of $CH_3NH_3PbI_3$ was prepared by mixing $PbI_2$ and $CH_3NH_3I$ in an equimolar ratio in γ-Butyrol lactone (GBL). To coat the substrate, the solution was spin-coated in two stages; 5 sec at 500 RPM and then at 1500 RPM for 30 sec. The substrate was then heated on a hot plate set at 100°C for 45 min, after which the substrate turned deep dark brown in color.

To finish the device fabrication, a 100 μL hole conductor solution (84 mg spiro-MeOTAD in 1 ml chlorobenzene, mixed with 7 μL of tert-butylpyridine and 15 μL of 170 mg/ml LiTFSI, bis(trifluoromethane)sulfonamide, in acetonitrile), was applied by spin coating 5 sec at 500

RPM, then at 1500 RPM for 30 sec, and then 100 nm gold contacts were thermally evaporated on the back through a shadow mask with 0.24 cm$^2$ rectangular holes.

*Device characterization:* The J-V characteristics were measured with a Keithley 2400-LV SourceMeter and controlled with a Labview-based, in-house designed program. A solar simulator (ScienceTech SF-150) equipped with a 1.5AM filter and calibrated with a Si solar cell IXOLAR$^{TM}$ High Efficiency SolarBIT (IXYS XOB17-04x3) was used for illumination. The devices were characterized through a 0.16 cm$^2$ mask.

*Electron Beam-Induced Current (EBIC) Measurements:* After characterizing the devices in the solar simulator, the samples were cleaved by hand after scribing the glass on the back and immediately loaded into the scanning electron microscope (SEM) vacuum chamber for EBIC measurements. SEM images were taken on a Zeiss SUPRA high-resolution microscope equipped with a Specimen Current Amplifier (GW electronics Inc. Type 31). The gold back contact was connected to the sample holder and, through it, to the preamplifier by a small micromanipulator to enable measuring the sample current. Using the Grün formula, we calculated a penetration depth of ca. 20 nm under working conditions[16] (1.5 keV; working distance of ca. 5 mm and density of the perovskite taken as ca. 4 g/mL), suggesting a lateral EBIC resolution of the same order of magnitude. Images were processed using the Fiji package of ImageJ 1.48.

*Kelvin probe force microscopy (KPFM) measurements:* KPFM is an atomic force microscopy (AFM) technique that enables to measure the sample surface potential with nanometer spatial resolution and mV sensitivity. Amplitude modulation (AM) KPFM measurements were carried out with a commercial AFM (Dimension Edge, Bruker Inc.) inside a N$_2$-filled glove box with < 1 ppm H$_2$O. The contact potential difference (CPD) was measured simultaneously with the topographic signal at an effective tip sample distance of 5 nm during scanning. The topographic height was obtained by maintaining the amplitude of the first cantilever resonance ($f_{1st}$ = 83 kHz) at a predefined amplitude set point of approximately 7 nm. The CPD was determined by compensating the ac component of the electrostatic force, $F_\omega$, at angular frequency ω with an applied dc voltage (= CPD) in a feedback control loop. To separate topographic from CPD signal, increase the sensitivity, and minimizing probe convolution effects, the ac electrostatic force component was generated at the second resonance, $f_{2nd}$ = 520 kHz, of the cantilever by applying an ac voltage of 500 mV. A Pt/Ir coated conductive probe (ATEC, Nanosensors) with a 40° tilted tip of 20 $\mu$m length was used for KPFM.

**Supplementary Information.**

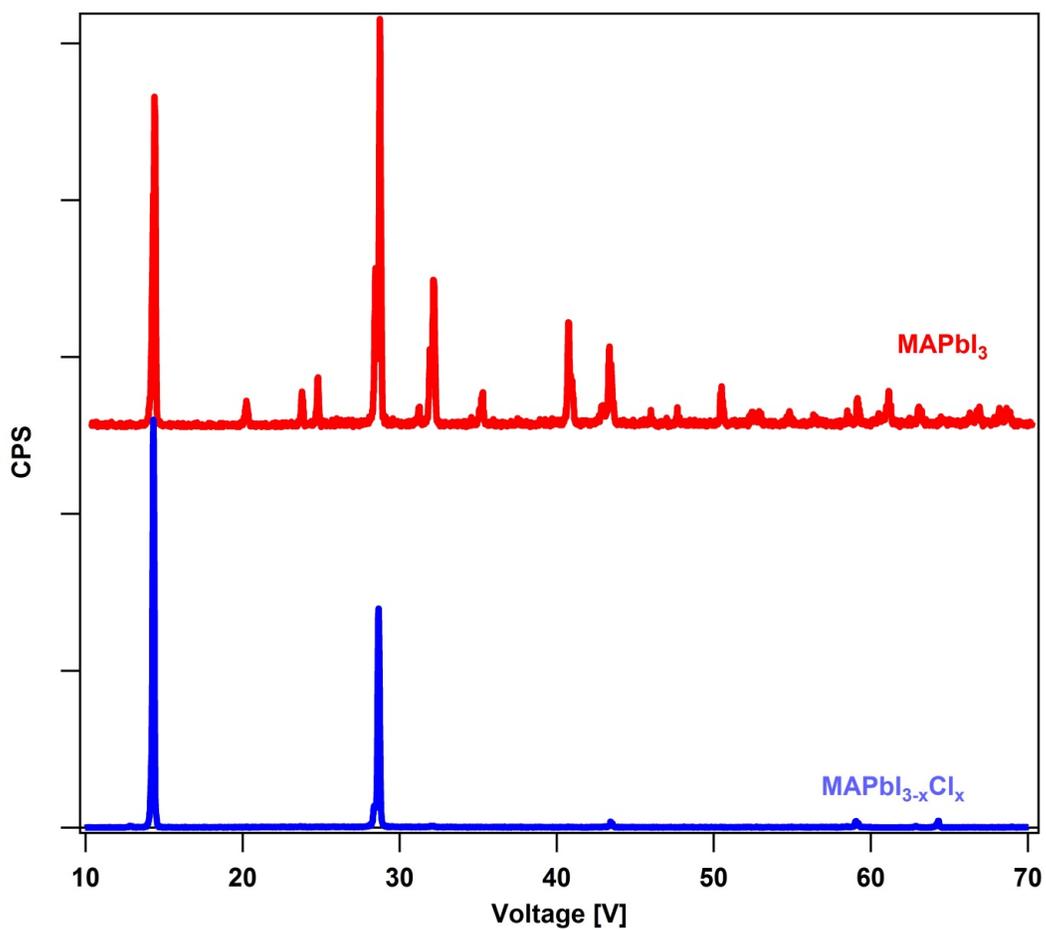

**Figure S1.** An X-ray diffraction of $CH_3NH_3PbI_3$ (red, top line) and $CH_3NH_3PbI_{3-x}Cl_x$ spin coated on glass substrate from DMF solution and annealed at 120C for 45 min in air.

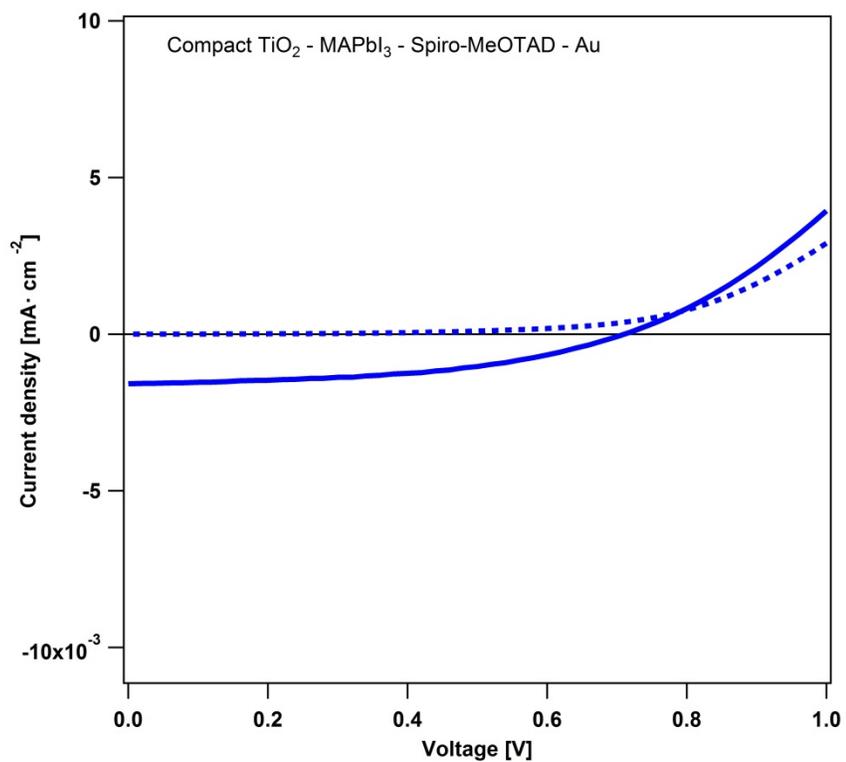

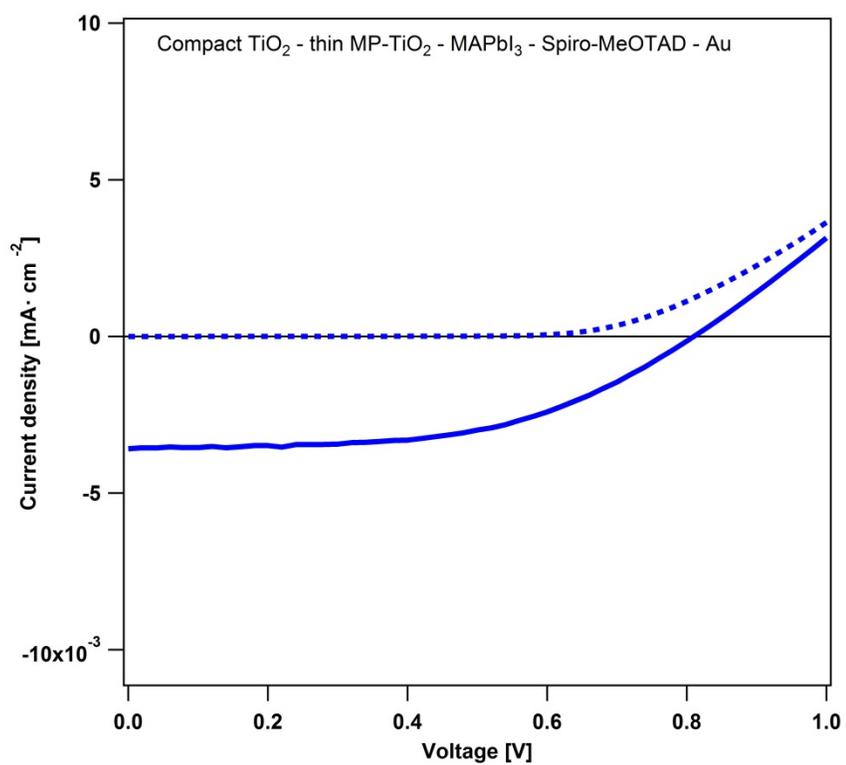

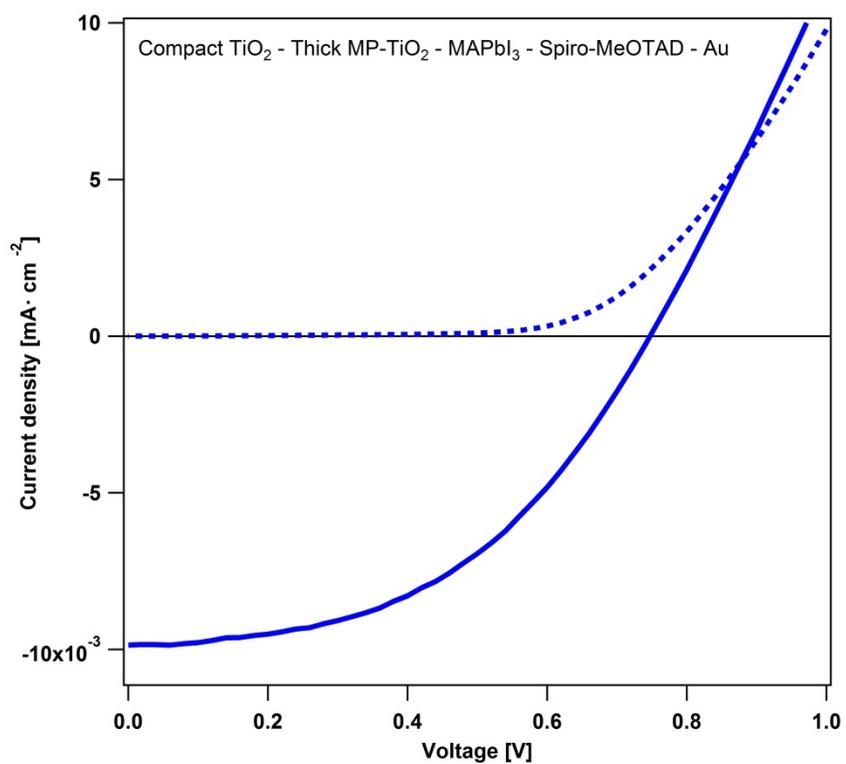

**Figure S2.** J-V characteristics of three cell architectures which were used for EBIC measurements a) FTO\compact TiO$_2$\MAPbI$_3$\Au b) FTO\compact TiO$_2$\thin mesoporous TiO$_2$\ MAPbI3\Spiro-MeOTAD\Au c) FTO\compact TiO$_2$\thick mesoporous TiO$_2$\ MAPbI3\Spiro-MeOTAD\Au